\documentclass[aps,prd,floats,nofootinbib,preprintnumbers]{revtex4}
\usepackage{epsfig}

\def\beq{\begin{equation}}
\def\eeq{\end{equation}}
\def\bea{\begin{eqnarray}}
\def\eea{\end{eqnarray}}

\date{\today}

\begin{document}

\preprint{NUHEP-TH/05-08}

\title{Low-Energy Neutrino Majorana Phases and Charged-Lepton Electric Dipole Moments}

\author{Andr\'e de Gouv\^ea}
\affiliation{Northwestern University, Department of Physics \& Astronomy, 2145 Sheridan Road, Evanston, IL~60208, USA}

\author{Shrihari Gopalakrishna}
\affiliation{Northwestern University, Department of Physics \& Astronomy, 2145 Sheridan Road, Evanston, IL~60208, USA}

\begin{abstract}
If the neutrinos are Majorana fermions, there are at least three new, potentially observable CP-odd phases that parameterize CP-invariance violating phenomena. We currently have no information regarding any of them and know that two out of the three, the so-called Majorana phases, are very hard to access experimentally. Here, we discuss the effect of Majorana phases on charged-lepton electric dipole moments (EDM), and explicitly show that neutrino Majorana phases induce EDMs even in the absence of other sources of CP-invariance violation. We also argue that while the Majorana neutrino contribution to EDMs is tiny, there is one generic ultraviolet completion to the standard model plus massive Majorana neutrinos --- the standard model plus a triplet Higgs boson --- that leads to significantly enhanced contributions which are still proportional to the low-energy neutrino Majorana phases. If this particular scenario is realized in nature, it seems possible, at least in principle, to measure the Majorana phases by precisely measuring charged-lepton EDMs.  

\end{abstract}
\maketitle

\setcounter{equation}{0}
\section{Introduction}
\label{intro}
 
 In the old standard model, all CP-invariance violating phenomena are parameterized by one dimensionless constant: the CP-odd phase factor $\delta_{CKM}$ contained in the quark mixing matrix. The existence of new sources of CP-invariance violation, however, is widely expected, thanks to the discovery of neutrino masses and lepton mixing \cite{nu_review}.    
  
If neutrinos are, just like the quarks and the charged-leptons, Dirac fermions, the three-by-three lepton mixing matrix is also expected to contain one potentially physical CP-odd phase factor $\delta$, which leads, for example, to different oscillation probabilities for CP-conjugated channels: $P(\nu_{\alpha}\to\nu_{\beta})\neq P(\bar{\nu}_{\alpha}\to\bar{\nu}_{\beta})$, $\alpha,\beta=e,\mu,\tau$. 
The experimental search for this kind of so-called leptonic CP-invariance violation is among the most important goals of the next-generation of high energy physics experiments. 
 
 If the neutrinos are Majorana fermions, leptonic CP-invariance violation can be much richer. In this case, the leptonic mixing matrix is parameterized by three potentially physical CP-odd phases $\delta,\eta,\zeta$. These will be properly defined in Sec.~\ref{sec:phases}. Physical phenomena predominantly sensitive to the so-called Majorana phases $\eta,\zeta$ are rare and hard to come by. The reason is quite simple --- Majorana phases are only physical in the advent of nonzero neutrino masses, and hence the amplitude for any process that involves a Majorana phase is directly proportional to the neutrino masses $m_i$, and these are typically much (much) smaller than the energies involved in particle physics processes: $A_{\rm Maj}\propto (m_i/E)^n$, where $n=1,2,3,\ldots$ 
 
Experimentally, the most sensitive probe of the effects of (a linear combination of) Majorana phases is the rate of nuclear neutrinoless double-beta decay. It is curious, however, that the Majorana phases affect the neutrinoless double-beta decay rate in a (predominantly) CP-even way \cite{cp_maj}. This does not mean, of course, that Majorana phases do not lead to CP-invariance violation. In \cite{cp_maj}, for example, several probes of CP-odd effects mediated by neutrino Majorana phases are discussed. There, only phenomena where lepton number is violated are discussed. 

In this paper, we study the effect of the neutrino Majorana phases in a lepton-number conserving but purely CP-invariance violating observable: the electric dipole moment (EDM) of charged leptons. Naively, one may imagine that, since neutrino Majorana phases are only physical if the neutrinos are Majorana fermions and lepton number is not exactly conserved, lepton-number conserving observables should be oblivious to them. This is known not to be the case \cite{nieves_pal}, and we explicitly show that even in the hypothetical case that neutrino Majorana phases are the only source of CP-invariance violation, charged leptons are expected to have nonzero EDMs.  

In Sec.~\ref{sec:nu_edm}, we review the two-loop massive Majorana neutrino contribution to charged lepton EDMs, and explicitly show its dependency on the Majorana phases. The massive neutrino contribution is tiny, even when compared with the (higher order) quark contribution, proportional to $\delta_{CKM}$. Nonetheless, it is guaranteed to be present and, in principle,\footnote{In practice, this is completely out of the question, as will become clear in Sec.~\ref{sec:nu_edm}.} precise measurements of charged lepton EDMs could be used to determine the value of particular linear combinations of the neutrino Majorana phases.

It is well known that the standard model plus massive Majorana neutrinos is only an effective theory. Hence, one needs to add new degrees of freedom at some ultraviolet scale $\Lambda$ in order to properly ``explain'' the origin of the neutrino masses.  Regardless of the mechanism behind neutrino masses, the new ultraviolet physics will also contribute ``directly'' to charged-lepton EDMs, often masquerading the effect of neutrino Majorana phases. This is particularly relevant  because of the absurdly suppressed nature of the massive neutrino contribution. In Sec.~\ref{sec:triplet}, we discuss one specific ultraviolet completion of the standard model plus massive Majorana neutrinos: the addition of an $SU(2)_L$ triplet Higgs boson. We show that the contributions of the triplet Higgs boson to charged lepton EDMs can be many orders of magnitude larger than the neutrino contributions. Nonetheless, these contributions are determined in terms of the ``low-energy'' lepton mixing angles and CP-odd parameters, including the Majorana phases. We contrast this behavior with, for example, that of the more popular type-I seesaw scenario \cite{seesaw}, recently examined in \cite{Archambault:2004td}.

Sec.~\ref{conclusions} contains some concluding remarks. Among other thoughts, we raise a rather peculiar hypothetical possibility:  in a world where there is only leptonic Majorana CP-invariance violation ($\delta_{CKM}=0$, $|U_{e3}|=0$, {\it cf.} Sec.~\ref{sec:phases}) it seems possible to determine the nature of the neutrino (Majorana fermion versus Dirac fermion) by measuring lepton-number conserving, but CP-invariance violating, observables. 
 
\section{Lepton Mixing and CP-odd Phases}
\label{sec:phases} 
After electroweak symmetry breaking, the leptonic sector of the standard model is parameterized by three charged lepton masses ($m_{\alpha}=m_e, m_{\mu}, m_{\tau}$), three neutrino masses ($m_i=m_1,m_2,m_3$), and a three-by-three unitary matrix $U$. We first choose to work in the basis where the weak charged current and the charged lepton masses are diagonal. In this so-called interaction (or weak, or flavor) basis,   

\beq
{\cal L}_{mass}^\nu = -\frac{1}{2}\overline{\nu^c}_\alpha \bar{m}_{\alpha\beta} \nu_\beta + h.c. \ ,  
\label{eq:Lnumeff}
\eeq
where $\nu^c \equiv (-i\gamma^2)\nu^*$ is the charge-conjugated field, $\bar{m}$ is a 
complex symmetric Majorana mass matrix and $\alpha,\beta=e,\mu,\tau$. Such a Majorana mass violates lepton number ($L_\#$) by two units. 

We can redefine the neutrino fields via
$\nu_\alpha = U_{\alpha i} \nu_i$, where the unitary matrix $U$ is chosen such that
$U^T \bar{m} U \equiv m$ is diagonal. We further choose the nonzero entries of $m=m_1,m_2,m_3$ to be nonnegative and real. $\nu_i$ are referred to as the neutrino mass eigenstates with masses $m_i$, $i=1,2,3$.\footnote{We will order the neutrino masses in the following way: $m_1<m_2$, $|m_3^2-m_1^2|>m^2_2-m_1^2$. $m_3>m_2$ characterizes a normal mass hierarchy, while $m_3<m_1$ an inverted one. For details see, for example, \cite{nu_review}.} Finally we choose the so-called charge-conjugation phase factors for the neutrinos to be equal to one \cite{cc_phases}.\footnote{If the neutrino are Majorana fermions, their fields obey the Majorana condition: $\nu_i=e^{\alpha_i}\nu_i^c$, for $i=1,2,3$. The charge-conjugation phases $\alpha_i$ are real. We choose $\alpha_i=0$, $\forall i$.} 

 In this new basis (the mass basis), the charged current interactions are described by
\begin{equation}
{\cal L}_{CC} = \frac{g}{\sqrt{2}} \overline{\ell}_\alpha \gamma^\mu P_L U_{\alpha i} \nu_i W_\mu^- + h.c.,
\end{equation}
where $\ell_{\alpha}$ are the charged lepton left-chiral fields, $e,\mu,\tau$ and $P_L=\frac{1}{2}(1-\gamma_5)$.

The unitary matrix $U$ can be parameterized by 
\begin{eqnarray}
&U=U^{\prime}P, \\
&U^{\prime}=\pmatrix{\cos\theta_{13}\cos\theta_{12} & \cos\theta_{13}\sin\theta_{12} & \sin\theta_{13}e^{-i\delta} \cr -\cos\theta_{23}\sin{\theta_{12}}-\sin\theta_{23}\cos{\theta_{12}}\sin\theta_{13}e^{i\delta} & \cos\theta_{23}\cos{\theta_{12}} -\sin\theta_{23}\sin{\theta_{12}}\sin\theta_{13}e^{i\delta} & \cos\theta_{13}\sin\theta_{23} \cr \sin\theta_{23}\sin{\theta_{12}}-\cos\theta_{23}\cos{\theta_{12}}\sin\theta_{13}e^{i\delta} & -\sin\theta_{23}\cos{\theta_{12}}-\cos\theta_{23}\sin{\theta_{12}}\sin\theta_{13}e^{i\delta} & \cos\theta_{13}\cos\theta_{23}}, \\
&P=\pmatrix{e^{i\eta/2} & \ & \cr \ & e^{i\zeta/2} & \ \cr \ & \ & 1} \ , 
\end{eqnarray}
where $\theta_{12},\theta_{23},\theta_{13}$ are the leptonic mixing angles, $\delta$ will be referred to as the `Dirac CP-odd phase,' and $\eta,\zeta$ will be referred to as `Majorana CP-odd phases.' 

The elements of $U^{\prime}$ are probed by neutrino$\leftrightarrow$neutrino oscillation experiments \cite{cp_maj}, together with the neutrino mass-squared differences, $\Delta m^2_{12}$ and $\Delta m^2_{13}$. Neutrino oscillation experiments have measured $|\Delta m^2_{13}|\sim 2\times 10^{-3}$~eV$^2$, $\Delta m^2_{12}\sim 8\times 10^{-5}$~eV$^2$, $\sin^2\theta_{23}\sim 0.5$, and $\sin^2\theta_{12}\sim 0.3$ \cite{global_anal}. Furthermore, the current data constrain $\sin^2\theta_{13}\lesssim 0.04$, and we have no information concerning the CP-odd phase $\delta$. Note that $\delta$ is only a physical observable in the event that $\sin^2\theta_{13}\neq 0$ --- the CP-odd phase $\delta$ can be ``redefined away'' if $|U_{e3}|$ vanishes.

Experimentally, nothing is known about the elements of $P$. Indeed, we don't even know if they are physical observables or not! We do know, however, that only physical processes that vanish in the limit $m_i\to 0$, $\forall i$ are sensitive to $\eta,\zeta$. There are several ways of showing that. The easiest, perhaps, is to perform another weak basis change: $\nu_i\to\nu_i^{\prime}=P_{ii}\nu_i$. In this basis, 
\begin{equation}
{\cal L}_{CC} +{\cal L}^{\nu}_{mass}= \frac{g}{\sqrt{2}} \overline{\ell}_\alpha \gamma^\mu P_L U^{\prime}_{\alpha i} \nu^{\prime}_i W_\mu^- -\frac{1}{2}\overline{\nu^{\prime c}}_i \hat{m}_{ij} \nu^{\prime}_j +  h.c.,
\end{equation}
where $\hat{m}$ is a diagonal complex matrix whose elements are given by $m_iP_{ii}^{-2}$. Hence, the Majorana phases can also be interpreted as the (relative) phases among the (complex) neutrino masses. Given our current understanding of neutrino masses and mixing, if the neutrinos are Majorana fermions, at least one of the Majorana phases is guaranteed to be an observable. In the advent that the lightest neutrino mass is identically zero, one of Majorana phases can be ``rotated away.'' 

In summary, if the neutrinos are Majorana fermions, our current understanding of neutrino masses and mixing guarantees that there is at least one new observable CP-odd phase in the new standard model --- a Majorana phase. If $\theta_{13}\neq 0$, the Dirac phase $\delta$ is also a physical parameter, while if the lightest neutrino mass is not zero, a second Majorana phase is also physical.\footnote{It should be clear that  ``physical'' CP-odd phases can be very small or even vanish, in which case no CP-invariance violating effects can be observed.} We will refer to the (currently experimentally allowed) $|U_{e3}|=0$, $m_{\rm lightest}=0$ case as the minimal leptonic CP-invariance violating scenario, or ``minimal leptonic CP-violation.''

\setcounter{equation}{0}
\section{Massive Neutrino Contribution to Charged Lepton EDM's} 
\label{sec:nu_edm}

\begin{figure}
\begin{center}
\scalebox{1.1}{\includegraphics{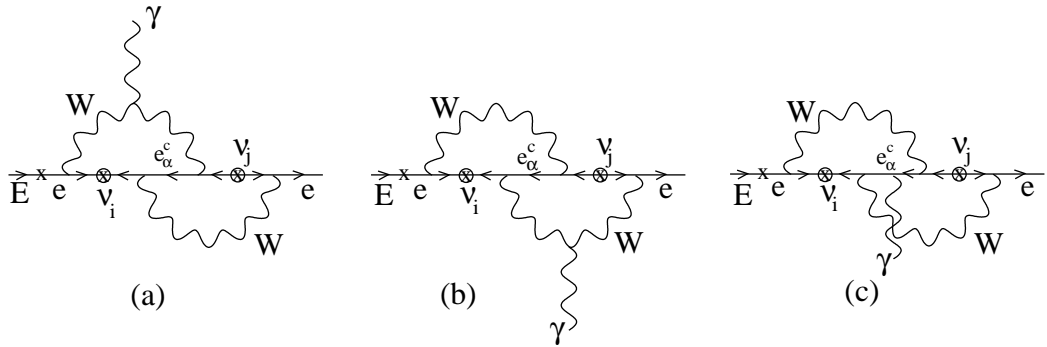}}
\caption{Electron EDM from Majorana neutrino and $W$ exchange. The arrows represent the flow of lepton number, $E$ ($e$) represents a right-handed (left-handed) electron, $\times$ represents a charged lepton mass insertion, and $\otimes$ represents a neutrino Majorana mass insertion. 
\label{fig:edm2nu}}
\end{center}
\end{figure}

Here, we would like to review that, if the neutrinos are Majorana fermions, a nonzero charged lepton EDM is induced at the two-loop level (see, for example, \cite{Ng:1995cs,Archambault:2004td}). The relevant two-loop diagrams that lead to a nonzero electron EDM are depicted in Fig.~\ref{fig:edm2nu}. Furthermore, we show that even if there is no ``Dirac'' CP-invariance violation ($|U_{e3}|=0$) and if the lightest neutrino mass vanishes (minimal leptonic CP-violation), the charged lepton EDM is sensitive to the only physical CP-odd phase in the lepton mixing matrix, at least in principle.

It is not difficult to qualitatively extract the electron EDM dependency on the neutrino masses and elements of the lepton mixing matrix. The diagrams in Fig.~\ref{fig:edm2nu} involve two Majorana mass insertions~\cite{Ng:1995cs,Archambault:2004td}, so that the imaginary part of the amplitude associated with diagram  Fig.~\ref{fig:edm2nu}(a) is given by
\beq
{\rm Im}\left[{\cal A}^{(a)}\right] = e \sum_{i,j,\alpha}\left(\frac{g}{\sqrt{2}}\right)^4 \Gamma^{e\alpha}_{ij} m_e m_i m_j 
  \left[ \frac{1}{M_W^4} \frac{1}{\left(16\pi^2\right)^2} f_2^{WW}\left(m_i^2,m_j^2,m_\alpha^2\right) \right] ,
\label{eq:aa}
\eeq
where
\beq
\Gamma^{e\alpha}_{ij} \equiv {\rm Im}\left( U^*_{ei} U^*_{\alpha i} U_{\alpha j} U_{ej} \right),
\label{eq:defGamma}
\eeq
and $f_2^{WW}$ is a dimensionless 2-loop function, given, for example, in \cite{Ng:1995cs}. The dependency on the leptonic CP-odd phases is contained in the $\Gamma^{e\alpha}_{ij}$ coefficients. These can be easily written in terms of phase reparameterization invariants, as discussed, for example, in \cite{nieves_pal}. Here, however, we find it more useful to stick to the parameterization presented in Sec.~\ref{sec:phases} in order to directly present the dependency on the Majorana phases, as will become clear shortly.

The nonzero contribution of $f_2^{WW}$ to ${\rm Im}\left[{\cal A}^{(a)}\right]$ is suppressed by powers of $m_{\alpha}^2/M^2_W$ and $m_i^2/M_W^2$. If one expands $f_2^{WW}$ in powers of $m_\alpha^2/M_W^2$, the zeroth-order term vanishes, since  $\sum_{\alpha}\Gamma^{e\alpha}_{ij}=0$. If one further expands in powers of $m_i^2/M_W^2$ and $m_j^2/M^2_W$ only the terms proportional to $m_{i,j}^2/M^2_W$ survive, given that $\sum_{ij}\Gamma^{e\alpha}_{ij}=0$. Finally, we also need $f_2^{WW}(m_i^2,m_j^2)\neq f_2^{WW}(m_j^2,m_i^2)$ in order to obtain a nonzero result. In summary, ${\rm Im}\left[{\cal A}^{(a)}\right] $ is severely ``GIM'' suppressed:
 \beq
{\rm Im}\left[{\cal A}^{(a)}\right] \propto \frac{(m_j^2 - m_i^2)}{M_W^2}\frac{m_{\alpha}^2}{M_W^2}.
\label{eq:mmlmwsup}
\eeq

It is easy to show that diagram Fig.~\ref{fig:edm2nu}(b), with $\nu_i \leftrightarrow \nu_j$, is identical to diagram Fig.~\ref{fig:edm2nu}(a) \cite{Ng:1995cs}. Hence, ${\cal A}^{(b)}$ is given by Eq.~(\ref{eq:aa}) with $i\leftrightarrow j$. Using $\Gamma^{e\alpha}_{ji}=-\Gamma^{e\alpha}_{ij}$, 
\beq
{\rm Im}\left[{\cal A}^{(a)+(b)}\right] = e \sum_{i,j,\alpha}\left(\frac{g}{\sqrt{2}}\right)^4 \Gamma^{e\alpha}_{ij} m_e m_i m_j 
  \left[ \frac{1}{M_W^4} \frac{1}{\left(16\pi^2\right)^2} \left(f_2^{WW}\left(m_i^2,m_j^2,m_\alpha^2\right)- f_2^{WW}\left(m_j^2,m_i^2,m_\alpha^2\right)\right)\right].
\eeq
Following the arguments in the previous paragraph, the contributions of Figs.~\ref{fig:edm2nu}(a) and (b) interfere constructively.

The amplitude associated with the diagram represented in Fig.~\ref{fig:edm2nu}(c) is also given by a variation of Eq.~(\ref{eq:aa}), with $f^{WW}_2$ replaced by a different dimensionless function $f_2^{\prime WW}$ of  the lepton masses. In this case, however, $f_2^{\prime WW}(m_i,m_j)=f_2^{\prime WW}(m_j,m_i)$, and the sum over $i$ and $j$ vanishes.
  
The two-loop neutrino contribution to the electron EDM $d_e$ is estimated to be 
\beq
d_e \sim \sum_{i<j,\alpha} e G_F^2 \Gamma^{e\alpha}_{ij} m_e m_i m_j \frac{(m_j^2 - m_i^2)}{M_W^2} \frac{m_\alpha^2}{M_W^2} \left[ \frac{1}{\left(16\pi^2\right)^2} \hat{f}_2^{WW}  \right].
\label{eq:deWW}
\eeq 
Since $m^2_{\tau}\gg m^2_{\mu}, m_e^2$, we concentrate on $\sum_{i<j} \Gamma^{e\tau}_{ij}m_i m_j$, and restrict our discussion to the minimal leptonic CP-violating scenario. We further assume that the neutrino mass hierarchy is inverted,\footnote{In the case of a normal hierarchy and minimal leptonic CP-violation, $\sum_{i<j}\Gamma^{e\alpha}_{ij}m_i m_j$ vanishes and there is no two-loop neutrino contribution to the electron EDM. There would be, of course, nonzero contributions to the muon and the tau EDMs.} so that 
\begin{eqnarray}
\sum_{i<j}  \Gamma^{e\tau}_{ij}m_i m_j&=&m_1m_2|U_{e1}||U_{\tau 1}||U_{e2}||U_{\tau 2}|{\rm Im}\left[-e^{i(\zeta-\eta)}\right], \\
&=&m_1m_2\cos^2\theta_{12}\sin^2\theta_{12}\sin^2\theta_{23}\sin(\eta-\zeta),
\end{eqnarray}
where, given an inverted mass hierarchy and $m_3=0$, $m_2\sim m_1=\sqrt{|\Delta m^2_{13}|}$. The electron EDM is sensitive to the only physical leptonic CP-odd phase, $\zeta-\eta$, and, as advertised, the effect is directly proportional to the neutrino masses. Indeed, the neutrino contribution to the leptonic EDM is suppressed by two Majorana neutrino mass insertions. This is guaranteed to happen, given the lepton-number conserving nature of the charged lepton EDM. In other words, one may interpret the Majorana neutrino phase contribution as a process that violates lepton number ``one minus one'' times \cite{nieves_pal,cp_maj}.

Given our current understanding of neutrino masses and leptonic mixing, the neutrino contribution to $d_e$ in the minimal leptonic CP-violation scenario is small beyond all reason: $d_e \lesssim 8\times 10^{-73}$~$e$-cm. This is to be compared with the contribution due to CP-invariance violation in the quark sector, which is estimated to contribute at the four-loop level: $d_e^{CKM}\lesssim 10^{-38}$~$e$-cm \cite{Ng:1995cs,Pospelov:1991zt}. There are also other purely leptonic contributions that do not depend on the nature of the neutrino (Dirac versus Majorana) and contribute at the three-loop level. We estimate these to add up to  $d_e^{\delta}\lesssim10^{-59}|U_{e3}|\sin\delta$~$e$-cm \cite{edm_reviews}, and hence vanish in the case of minimal leptonic CP-violation. 

We note that similar expressions can be obtained for the muon and tau EDMs. As is often the case, $d_{\alpha}/d_{\beta}$ is expected to be (assuming all nonzero $\Gamma^{\alpha\beta}_{ij}$ are of the same order of magnitude) of order $m_{\alpha}/m_{\beta}$.

\setcounter{footnote}{0}
\setcounter{equation}{0}
\section{Ultraviolet Finite Example -- Triplet Higgs Model}
\label{sec:triplet}

Similar to charged fermion masses, neutrino Majorana masses are not allowed by the $SU(2)_L\times U(1)_Y$ gauge symmetry of the standard model. Furthermore, given the old standard model particle content, neutrino masses are constrained to be zero even after electroweak symmetry is spontaneously broken. Of course, neutrino Majorana masses are present if one interprets the SM as an effective theory, valid below some energy scale $\Lambda$ \cite{so_what}. In this case, higher dimensional operators naively suppressed by powers of $\Lambda$ are to  be added to the Lagrangian:
\begin{equation}
{\cal L}_{5+} = -\frac{1}{2\Lambda}\overline{L^c}_\alpha \bar{\lambda}_{\alpha\beta} L_\beta HH - \frac{i}{\Lambda^2} H^{\dagger}\cdot\overline{L}_{\alpha}g_{\alpha\beta} \sigma_{\mu\nu}\gamma_5e_{\beta}F^{\mu\nu} + h.c. + \ldots \ ,  
\label{eq:5+}
\end{equation}
where $g_{\alpha\beta}$ are dimensionless constants, $L_{\alpha}$ ($e_{\beta}$) are lepton $SU(2)$ doublet (singlet) fields, and $H$ is the $SU(2)_L$ doublet Higgs boson field. After spontaneous electroweak symmetry breaking (when $H$ acquires a vacuum expectation value $v$), Eq.~(\ref{eq:5+}) leads to a neutrino mass matrix $\bar{m}=\bar\lambda v^2/\Lambda$ {\sl and} a charged lepton electric dipole moment matrix $d=gv/\Lambda^2$. More real progress can only be made once the renormalizable physics responsible for  Eq.~(\ref{eq:5+}) is specified. For example, it is not at all guaranteed that the suppression factor in the dimension five operator is the same as the suppression factor in the dimension six operator. Indeed, in several theories beyond the standard model, these are expected to be wildly different, given that the dimension five neutrino mass operator violates lepton number, while the EDM dimension six operator does not (for a list of examples, see, for example, \cite{edm_reviews}). Nonetheless, it is interesting to estimate what one expects in the case $\bar\lambda\sim 1$, $g\sim e m_{\ell}/v$.\footnote{We assume that the physics responsible for the EDM dimension six operator is proportional to the electron Yukawa coupling and the QED coupling $e$. This need not be the case.} Under these circumstances, $m\sim v^2/\Lambda$ and $d_{\alpha}\sim em_{\alpha}/\Lambda^2$. Hence,
\begin{equation}
d_e\sim eG_F^2m_e m_{\nu}^2=1.4\times 10^{-47}\left(\frac{m_{\nu}}{0.1~\rm eV}\right)^2~e\hbox{-cm},
\label{eq:naive_edm}
\end{equation}
where $m_{\nu}$ is a typical neutrino mass. This contribution, while very small, is many order of magnitude larger than the Majorana neutrino one obtained in the previous section, and hence dominates the ``new physics'' (which includes the generation of the dimension five, Majorana neutrino mass operator) contribution to the charged lepton EDMs. Furthermore, it should be readily seen that there is, {\it a priori}, no relation between the neutrino Majorana phases, contained in $\bar{\lambda}$, and the charged lepton EDMs.  

An explicit ultraviolet completion that leads to Eq.~(\ref{eq:5+}) is the well-known type-I seesaw mechanism \cite{seesaw}. In \cite{Archambault:2004td}, the seesaw contribution to charged lepton EDMs was computed. Modulo extraordinary circumstances (see \cite{Archambault:2004td} for details), the estimate quoted in Eq.~(\ref{eq:naive_edm}) was obtained, and there was no connection between the CP-odd parameters probed in the EDM experiment and low energy neutrino CP-odd phases.  

 A different origin for the neutrino masses is to postulate the existence of a new scalar field: an $SU(2)_L$ triplet Higgs boson $\xi$ with lepton number $L_{\#}=-2$. If such a field exists, it couples to the charged lepton doublets via the renormalizable interaction
 \begin{equation}
 {\cal L}_{L\xi L} =-\frac{\sqrt{2}}{2}\bar\kappa_{\alpha\beta} \overline{{L^c}^\alpha} \cdot \xi L^\beta + {\rm h.c.}.
 \end{equation}
If the neutral component of $\xi$ develops a vacuum expectation value $u$, the neutrinos acquire a mass
\beq
\bar{m}_{\alpha\beta} \equiv u \bar{\kappa}_{\alpha\beta}.
\label{eq:defmuk}
\eeq 
The SM plus the Higgs triplet Lagrangian is renormalizable, and one can compute its contribution to the charged lepton EDMs. Similar to the seesaw mechanism, we find that it can be much larger than the massive Majorana neutrino contribution. On the the other hand, the CP-odd parameters responsible for a nonzero EDM are trivially related to the low-energy leptonic Majorana phases. This is easy to see. The SM augmented by the triplet Higgs boson $\xi$ contains the same number of CP-odd phases as the SM after electroweak symmetry breaking augmented by Majorana neutrino masses. These phases are easiest to identify in the interaction basis --- they are encoded in the complex entries of $\bar{\kappa}$. From Eq.~(\ref{eq:defmuk}) it is trivial to see that the same unitary matrix that diagonalizes $\bar{m}$ will also diagonalize $\bar{\kappa}$: $U^T\bar{\kappa}U=\kappa$ where $U$ is the lepton mixing matrix discussed in Sec.~\ref{sec:phases} and $\kappa$ is a diagonal matrix of dimensionless couplings. 
The nonzero elements of $\kappa$ ($\kappa_{ii}\equiv\kappa_i$, $i=1,2,3$) are trivially related to the neutrino masses $m_i$:
\beq
m_i \equiv u \kappa_i \ ,
\eeq
where $\kappa_i$ are real and nonnegative.

It is possible to write down triplet Higgs boson models where $u$ is naturally small~\cite{Schechter:1980gr,Gelmini:1980re,Ma:1998dx} so that small neutrino masses are obtained for ``large'' $\kappa$ values and electroweak scale masses for the propagating scalar degrees of freedom. We present the details of such an extension of the SM in Appendix~\ref{appTriHiggs} and  concentrate here on presenting the Higgs triplet contributions to $d_e$. 

\begin{figure}
\begin{center}
\scalebox{1.1}{\includegraphics{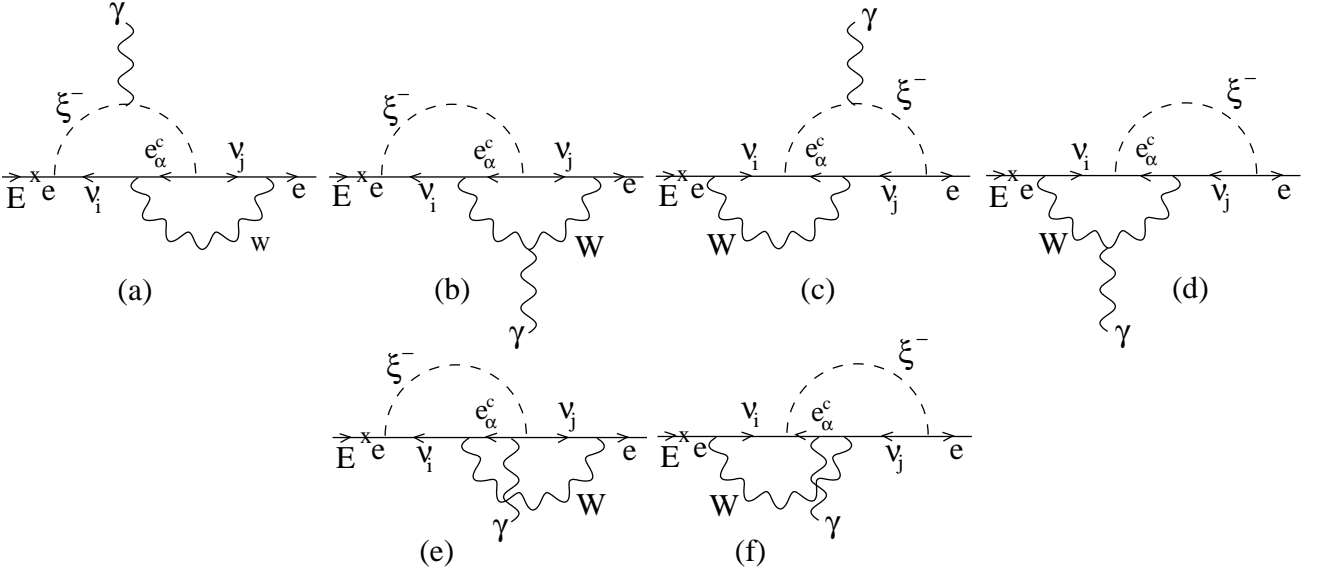}}
\caption{Set of two-loop diagrams in the triplet Higgs boson boson that contribute to the electron EDM. See Appendix~\ref{appTriHiggs} for details. The arrows represent the flow of lepton number, $E$ ($e$) represents a right-handed (left-handed) field, and $\times$ represents a charged lepton mass insertion.
\label{fig:edmWxi}}
\end{center}
\end{figure}

Fig.~\ref{fig:edmWxi} depicts two-loop contributions to the electron EDM arising from the exchange of $W$ and $\xi$.
Comparing these to the diagrams in Fig.~\ref{fig:edm2nu}, we note that one of the $W$ loops is replaced
by a $\xi$ loop. The main consequence of that is that  the diagrams in Fig.~\ref{fig:edmWxi} are no longer suppressed by Majorana neutrino mass insertions. Here, the role played by the $m_i$ is promptly played by the dimensionless couplings $\kappa_i$. 
The amplitude for diagram Fig.~\ref{fig:edmWxi}(a) is given as
\beq
{\rm Im}\left[{\cal A}^{(a)}\right] = \sum_{i,j,\alpha}e g^2 \kappa_i \kappa_j \Gamma^{e\alpha}_{ij} m_e \left[ \frac{1}{\tilde{M}^2} \frac{1}{\left(16\pi^2\right)^2} f_2^{\xi W}\left(m_i^2,m_j^2,m_\alpha^2\right) \right] \ , 
\eeq
where $\tilde{M}$ is symbolic for a combination of the heavy mass scales $M_W$ and $M$ (where $M$ 
is the triplet Higgs boson mass, defined in Eq.~(\ref{eq:Vxidagxi})), the triplet Higgs boson couplings
to fermions $\kappa_i$ are defined in Eq.~(\ref{eq:LxiLm}),  and $f_2^{\xi W}$ is again a dimensionless two-loop function of particle masses. As in the previous section, the sum over $\alpha,i,j$ vanishes if one ignores the $m_\alpha^2,m_i^2,m_j^2$ dependency of $f_2^{\xi W}$ given $\sum_\alpha\Gamma^{e\alpha}_{ij}=0$ and $\sum_{ij}\Gamma^{e\alpha}_{ij}=0$. One ends up, as before, with a large suppression factor, such that
\beq
{\rm Im}\left[{\cal A}^{(a)}\right] \propto \frac{(m_j^2 - m_i^2)}{\tilde{M}^2} \frac{m_\alpha^2}{\tilde{M}^2} \ . 
\eeq
Following the reasoning used in the previous section, ${\rm Im}\left[{\cal A}^{(a)}\right]={\rm Im}\left[{\cal A}^{(c)}\right]$, and the other two-loop diagrams in Fig.~\ref{fig:edm2nu} are expected to add similar contributions. These two-loop $W-\xi$ diagrams contribute to the electric dipole moment an amount  
\beq
d_e^{\xi W} \sim \sum_{i<j,\alpha} e g^2 \kappa_i \kappa_j \Gamma^{e\alpha}_{ij} m_e \frac{(m_j^2 - m_i^2)}{\tilde{M}^2} \frac{m_\alpha^2}{\tilde{M}^2} \left[ \frac{1}{\tilde{M}^2} \frac{1}{\left(16\pi^2\right)^2} \hat{f}_2^{\xi W} \right] \ . 
\label{eq:xi2loop}
\eeq
For order one values of $\kappa_i$ and $\hat{f}^{\xi W}_2$, Eq.~(\ref{eq:xi2loop}) is enhanced with respect to neutrino loop Eq.~(\ref{eq:deWW}) by a factor 
$(M_W^2/m_im_j)(M_W/\tilde{M})^6\sim 10^{18}$ for $\tilde{M}=1$~TeV, such that, in the case of minimal leptonic CP-violation and an inverted mass hierarchy, $d_e^{\xi W}\lesssim 10^{-55}$~$e$-cm. Note that this is still several orders of magnitude smaller than the expected four-loop $\delta_{CKM}$ contribution to the electron EDM.

\begin{figure}
\begin{center}
\scalebox{1.1}{\includegraphics{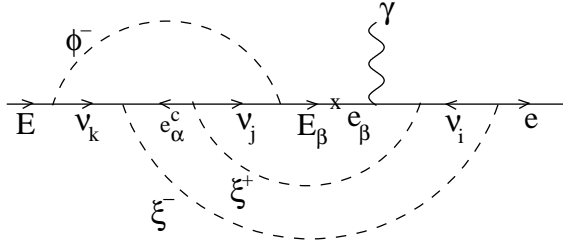}}
\caption{Three-loop diagrams in the triplet Higgs boson boson that contribute to the electron EDM. See Appendix~\ref{appTriHiggs} for details. The arrows represent the flow of lepton number, $E$ ($e$) represents a right-handed (left-handed) field, and $\times$ represents a charged lepton mass insertion. 
\label{fig:edmxiphi}}
\end{center}
\end{figure}

One should expect that, at higher loops, it is possible to obtain triplet Higgs boson contributions to the electron EDM which are not suppressed by $\Delta m^2_{ij}/\tilde{M}^2$ due to the fact that the triplet Higgs couplings to charged fermions are also not flavor universal (and are also given by $\bar{\kappa}$). For example, Fig.~\ref{fig:edmxiphi} depicts, in the 't\,Hooft-Feynman gauge, one three-loop contribution to the 
electron EDM due to the exchange of $\phi^-$ and $\xi^\pm$ (see Appendix~\ref{appTriHiggs} for details).
Defining 
\beq
\Gamma_{ijk}^{e\beta\alpha} = {\rm Im}\left[U_{ei} U^*_{\beta i} U_{\beta j} U_{\alpha j} U^*_{\alpha k} U^*_{ek} \right] \ ,
\eeq
we find the amplitude associated with Fig.~\ref{fig:edmxiphi} is
\beq
{\rm Im}\left[{\cal A}\right] = 4 e \sum_{i,j,k,\beta,\alpha}\frac{m_em_{\beta}^2}{v^2} \kappa_i^2 \kappa_j \kappa_k \Gamma_{ijk}^{e\beta\alpha} \left[ \frac{1}{\tilde{M}^2} \frac{1}{\left( 16\pi^2 \right)^3} f_3^{\phi\xi\xi}\left( m_\alpha^2,m_i^2,m_j^2,m_k^2 \right) \right] \ ,
\eeq
where $f^{\phi\xi\xi}_3$ is a dimensionless 3-loop function. As was the case of $\Gamma^{\alpha\beta}_{ij}$, $\Gamma^{\alpha\beta\gamma}_{ijk}$ can be easily written in terms of phase reparameterization invariants, as defined, for example, in \cite{nieves_pal}.

If we ignore the $m_{\alpha}$ dependency of $f^{\phi\xi\xi}_3$, we can easily perform
\begin{equation}
\sum_{\alpha}\Gamma_{ijk}^{e\beta\alpha}={\rm Im}\left[U_{ei}U_{\beta i}^*U_{\beta j}U_{ej}^*\right],
\end{equation}
which is none other than the familiar Jarlskog invariant \cite{Jarlskog:1985ht}. It does not depend on the Majorana phases and vanishes in the minimal lepton CP-violation scenario and, as we are concentrating on the impact of Majorana phases in general and the minimal leptonic CP-violation scenario in particular, we are left with
\beq
d_e^{\phi\xi\xi} \sim e \sum_{i,j,k}G_Fm_em_{\beta}^2 \kappa_i^2 \kappa_j \kappa_k \Gamma_{ijk}^{e\beta\alpha} \frac{m_\alpha^2}{\tilde{M}^2} \left[ \frac{1}{\tilde{M}^2} \frac{1}{\left( 16\pi^2 \right)^3} \hat{f}_3^{\phi\xi\xi}  \right] \ .
\eeq
It seems that there are no other ``GIM'' suppression factors (say, $\Delta m^2_{ij}/M_W^2$). We verify this explicitly  by performing the $\sum_{ijk}$, and obtain a nonzero result even if we ignore the dependency of $\hat{f}_3^{\phi\xi\xi}$ on the neutrino masses . In the case of a normal mass hierarchy and minimal leptonic CP-violation ({\it i.e.} $|U_{e3}|=0$, $m_1=0\to\kappa_1=0$), 
\begin{eqnarray}
\sum_{i,j,k}\Gamma^{e\tau\tau}_{ijk}\kappa_i^2\kappa_j\kappa_k&=&\sum_{j=2,3}{\rm Im}\left[U_{e2} U^*_{\tau 2} U_{\tau j} U_{\tau j} U^*_{\tau 2} U^*_{e2} \right]\kappa_2^3\kappa_j={\rm Im}\left[U_{e2} U^*_{\tau 2} U_{\tau 3} U_{\tau 3} U^*_{\tau 2} U^*_{e2} \right]\kappa_2^3\kappa_3, \\
&=&-\kappa_2^3\kappa_3\cos^2\theta_{12}\sin^2\theta_{12}\cos^2\theta_{23}\sin^2\theta_{23}\sin{\zeta},
\end{eqnarray}
where $\kappa_2=\kappa_3\sqrt{(\Delta m^2_{12}/\Delta m^2_{13})}\sim 0.2\kappa_3$.
As in the case of the light neutrino contribution computed in the previous section, $d_e^{\phi\xi\xi}$ is directly proportional to the sine of the only physical leptonic CP-odd phase ($\zeta$, in the case of a normal neutrino mass hierarchy and minimal leptonic CP-violation). From this, and assuming $\tilde{M} = 1~{\rm TeV}$, $\hat{f}_3^{\phi\xi\xi}\sim O(1)$, we estimate 
$d_e \lesssim 10^{-38}$~$e$-cm, which is of the same order of magnitude as the $\delta_{CKM}$ contribution. 

We consider $d_e^{\phi\xi\xi}$ as an estimate for the three-loop contribution of the triplet Higgs boson model described in Appendix \ref{appTriHiggs} to the electron EDM with reserved optimism. We searched for other ``large'' three-loop contributions to the electron EDM,  and Fig.~\ref{fig:edmxiphi} proved to be the most interesting one.  However, we have not (by any stretch of the imagination) performed the computation of the entire three-loop contribution. There remains the possibility that, once all three-loop diagrams are indeed included, extra suppression factors (or even exact cancellations) can appear. Nonetheless, the overarching message we wish to convey is robust and clear: triplet Higgs boson contributions to charged lepton EDMs are, if the triplet Higgs boson masses are of order the electroweak scale, many orders of magnitude larger than the neutrino contributions. It is also conceivable that the triplet Higgs boson contributions are the dominant source of charge lepton EDMs. Most importantly for us here is the fact that these contributions are proportional to the low-energy neutrino Majorana phases. Once all triplet Higgs boson masses and neutrino masses are known, a measurement of the charged lepton EDM can be translated, at least in principle, into a measurement of the neutrino Majorana phases.

\setcounter{equation}{0}
\section{Conclusions}
\label{conclusions}

With the discovery of neutrino masses and lepton mixing comes the expectation that there are CP-invariance violating phenomena which are not parameterized by the CP-odd phase $\delta_{CKM}$ of the quark mixing matrix. If the neutrinos are Dirac fermions, one extra observable CP-odd phase $\delta$, naively unrelated to $\delta_{CKM}$, is guaranteed to be present in the lepton mixing matrix as long as $|U_{e3}|\neq0$. If the neutrinos are Majorana fermions, two more phases $\eta$ and $\zeta$, known as Majorana phases, may also be potentially observable, as long as the lightest neutrino mass is not zero. It is curious to realize that, if the neutrinos are Majorana fermions, at least one\footnote{Two out of three CP-odd phases are unphysical if the lightest neutrino mass vanishes {\sl and} if $|U_{e3}|=0$. Both possibilities are perfectly allowed experimentally, and the scenario where both are realized is referred to here as ``minimal leptonic CP-violation.''} new CP-odd phase, naively unrelated to $\delta_{CKM}$, is already guaranteed to be a physical observable. The search for physical effects (especially CP-odd effects) mediated by these new CP-invariance violating parameters is among the highest priorities of fundamental particle physics research.

Here, we have reviewed that nonzero EDMs for the charged leptons are induced at the two-loop level in the standard model (SM) plus massive Majorana neutrinos. We then showed explicitly that these EDMs are sensitive to the Majorana phases even in the case of minimal leptonic CP-violation. Charged lepton EDMs qualify, therefore, as another CP-violating observable sensitive to the neutrino Majorana phases (see \cite{cp_maj,nieves_pal} for a discussion of other such observables).

There are (at least) two problems with exploring, in practice, charged lepton EDMs as means for measuring the leptonic CP-violating phases. One is that the purely leptonic contribution to the EDMs is tiny --- many orders of magnitude smaller than the four-loop quark contribution, proportional to $\delta_{CKM}$. The other is that physics beyond the SM will also contribute to  EDMs. Indeed, the new physics responsible for neutrino masses, guaranteed to be there, can, in general, contribute to the charged-lepton EDMs much more (by many orders of magnitude) than the massive Majorana neutrinos. Furthermore, even these new physics contributions are in general proportional to ``ultraviolet'' CP-odd parameters, such that the dependency on the low-energy neutrino Majorana phases is often lost. This is the case, for example, of the type-I seesaw mechanism \cite{Archambault:2004td}.

We showed that both problems raised above are absent (at some level) if the Majorana neutrino masses arise as a consequence of the existence of an $SU(2)_L$ triplet Higgs boson with $L_{\#}=-2$. Under the right circumstances, higher order corrections associated with the triplet Higgs bosons contribute to the charged lepton EDM as much as (and, perhaps more than) the quarks. Nonetheless, we show that these contributions are still proportional to the low-energy leptonic CP-odd phases, including the Majorana phases. If this is indeed the mechanism responsible for massive Majorana neutrinos, it may turn out that the dominant contribution to charged lepton EDMs is due to a neutrino Majorana phase. 

Even in the most optimistic scenario, we are still far from exploring neutrino Majorana phases via EDM searches. The current bound on $d_e=(6.9\pm 7.4)\times 10^{-28}~e$-cm \cite{pdg} is still some ten orders of magnitude larger than our largest estimates. In the future, significant improvements are expected, and one can hope to be sensitive to $d_e>10^{-31}~e$-cm \cite{demille}. This is still, alas, orders of magnitude away from the contributions discussed here. Finally, one must also contend with the fact that the electron used in EDM measurements is part of an atomic system, and that there are $\delta_{CKM}$ effects due to electron--nucleus interactions that also contribute to the atomic EDM, which is the experimental observable. See \cite{edm_reviews} and references therein for more details.  

We would like to conclude with a  ``theoretical'' question. In a hypothetical universe where $\delta_{CKM}=0$ and the minimal leptonic CP-violating scenario is realized, all CP-violating phenomena are parameterized by one single CP-odd phase --- a Majorana phase. Majorana phases, on the other hand, are only physical if the neutrinos are Majorana fermions, indicating that if the neutrinos were Dirac fermions, CP-invariance would be exact. It seems, therefore, possible to establish, in this hypothetical universe, that the neutrinos are Majorana fermions by determing that, say, the electron EDM is nonzero and hence CP-invariance is not conserved. The problem is that the electron EDM is not a lepton number violating observable. Is it really possible, at least in principle, to establish that the neutrinos are Majorana fermions without directly observing that lepton number is not a conserved quantum number?

\section*{Acknowledgments}

AdG thanks Boris Kayser for useful conversations, and SG thanks Alexey Pronin and Tatsu Takeuchi for discussions on two-point functions. We thank Bill Marciano and Tom Rizzo for suggesting M\o ller scattering and muonium-antimuonium oscillation as probes of the triplet Higgs model, and we are grateful to Maxim Pospelov for useful suggestions and comments on the manuscript. This work is sponsored in part by the US Department of Energy Contract DE-FG02-91ER40684.

\appendix
\section{Triplet Higgs Model Lagrangian and Feynman Rules}
\label{appTriHiggs}

We add to the SM a complex Higgs triplet~\cite{Schechter:1980gr,Gelmini:1980re,Ma:1998dx} which
we denote as $\xi = (\xi^{++}\ \xi^+\ \xi^0)^T$. Using the SU(2) generators $T^a = \sigma^a/2$, 
where $\sigma^a$ are the $2\times 2$ Pauli matrices, we can alternatively define 
\beq
\xi \equiv \xi^a T^a = \frac{1}{\sqrt{2}} \pmatrix{\xi^+/\sqrt{2} & \xi^{++} \cr \xi^0 & -\xi^+/\sqrt{2}} \ .   
\eeq
We add to the SM Lagrangian ${\cal L}_{SM}$ terms that couple $\xi$ to the SM fields as follows:
\bea
{\cal L} &=& {\cal L}_{SM} + {\cal L}_{\xi K.E.} + {\cal L}_{L\xi L} - V_\mu - V_{\xi^\dagger\xi} \ , \\
{\cal L}_{\xi K.E.} &=& 2 Tr\left[ (D^\mu\xi)^\dagger D_\mu\xi \right] \ , \\
{\cal L}_{L\xi L} &=& -\frac{\sqrt{2}}{2}\bar\kappa_{\alpha\beta} \overline{{L^c}^\alpha} \cdot \xi L^\beta + {\rm h.c.} \ , \label{eq:LxiL} \\
{\cal L}_{SM} &\supset& - V_{SM} = - m_0^2 - \frac{1}{2} \lambda_1 (H^\dagger H)^2 \ , \\
V_\mu &=& - \sqrt{2}\mu H^T \cdot \xi^\dagger H + {\rm h.c.} \ , \label{eq:Vmu} \\
V_{\xi^\dagger \xi} &=& 2 M^2 {\rm Tr}\left[ \xi^\dagger \xi \right] + 2 \lambda_2 \left\{ Tr\left[ \xi^\dagger \xi  \right] \right\}^2 + 2 \lambda_3 (H^\dagger H) Tr\left[ \xi^\dagger \xi \right] + 2 \bar{\lambda}_3 (H^\dagger T^a H) Tr\left[ \xi^\dagger T^a \xi \right] \ , \label{eq:Vxidagxi}   
\eea
where the dot denotes the antisymmetric $SU(2)_L$ product, $\alpha,\beta$ 
represent generation indices, $\bar\kappa$ is a complex symmetric matrix, and $H = (\phi^+ \ \phi^0)^T$
is the SM Higgs doublet. 
From the Lagrangian we see that $\xi$ has hypercharge $Y(\xi) = +1$, and we assign it lepton number 
$L_\# (\xi) = -2$. 

We note that the only term that explicitly breaks $L_\#$ is $V_\mu$. 
When $\mu = 0$, $L_\#$ is a good symmetry of the Lagrangian, and 
if $L_\#$ were then spontaneously broken by $\left< \xi^0 \right> \equiv u \neq 0$, 
we would obtain a massless ``majoron'' field, which leads to phenomenological problems (see, for example, \cite{Ma:1998dx}). 
This is avoided when $L_\#$ is explicitly broken by $\mu \neq 0$, and the would-be 
majoron mass is then of the order of $M$~\cite{Ma:1998dx}.
In this work, we consider the situation where $L_\#$ is not broken spontaneously
but is broken explicitly by a nonzero $\mu$. 

Minimizing the potential $V_{SM} + V_\mu + V_{\xi^\dagger\xi}$, and defining $\left< \xi^0 \right> \equiv u$ and $\left< \phi^0 \right> \equiv v$, we get 
\bea
u &=& -\frac{\mu v^2}{M^2} \frac{1}{\left[ 1 + (\lambda_3 + \bar{\lambda}_3)\frac{v^2}{M^2} \right]} \ \ \ \ ({\rm for}\ u \ll M) \ , \label{eq:umuM} \\
v &=& \sqrt{-\frac{m^2}{\lambda_1}-\frac{2\mu u}{\lambda_1}} \ \ \ \ ({\rm for}\ u \ll v)  \ .
\label{eq:uvVEV}
\eea 
As in the old SM, $v$ defines the scale of electroweak symmetry breaking.

A small neutrino mass can be obtained for order one values of $\kappa$ if $u\ll v$. For example, naturally small values of $u$ can be explained by a large mass scale $M$ of the 
triplet Higgs owing to the (type-II) seesaw relation in Eq.~(\ref{eq:umuM}). For example,
$\mu \sim v$ and $M\sim 10^8~{\rm GeV}$ leads to $u\sim 0.1~{\rm eV}$, as desired. However, if
such a situation is realized in nature, there would be no other probes of the triplet Higgs sector
at accessible energies other than the effective Majorana mass in Eq.~(\ref{eq:Lnumeff}). 

Another possibility (the one in which we are interested here) is that $M\sim v$ and $\mu\sim 0.1$~eV, and, from Eq.~(\ref{eq:umuM}),
$u\sim 0.1~{\rm eV}$, as required by the smallness of the neutrino masses. Unlike the case considered in the last paragraph, 
however, there is no ``explanation'' for why $\mu$ is so small, and we lose the familiar seesaw relation for $u$. While this is true, we point out that $u$ is proportional to $\mu$ and, if we set $\mu=0$,
we recover $L_\#$. Thus, a tiny $\mu$ is natural in the 't\,Hooft sense. For such an explanation
to be complete, it might be desirable to extend this model in order to incorporate a dynamical 
mechanism for generating such a tiny $\mu$. Compared to the previous case, this has the 
appeal that one could directly or indirectly probe the existence of the triplet Higgs boson. 
One such indirect observable is the lepton EDM, discussed in the main body of this paper. Others are briefly discussed below.

Expanding the $SU(2)_L$ structure of Eq.~(\ref{eq:LxiL})
\beq
{\cal L}_{L\xi L} = \frac{1}{2}\left(-\xi^0 \overline{\nu^c} \bar\kappa \nu + \frac{\xi^+}{\sqrt{2}} \left( \overline{e^c} \bar\kappa \nu + \overline{\nu^c} \bar\kappa e \right)  + \xi^{++} \overline{e^c} \bar\kappa e\right)  + {\rm h.c.} \ .
\label{eq:LxiLexp}
\eeq
The first term in Eq.~(\ref{eq:LxiLexp}), with $\left< \xi^0 \right> = u$, leads to a 
Majorana neutrino mass of the form shown in Eqs.~(\ref{eq:Lnumeff}), where  $\bar{m}$ is defined as in Eq.~(\ref{eq:defmuk}). 
Writing Eq.~(\ref{eq:LxiLexp}) in the mass basis we get
\bea
{\cal L}_{L\xi L} = \frac{1}{2}\left(-\xi^0 \overline{\nu^{c}} \kappa \nu + \sqrt{2} \xi^+ \overline{\nu^{ c}} \kappa U^\dagger e + \xi^{++} \overline{e^c} U^* \kappa U^\dagger e\right)  + {\rm h.c.} \ . 
\label{eq:LxiLm}
\eea

\begin{figure}
\begin{center}
\scalebox{1.1}{\includegraphics{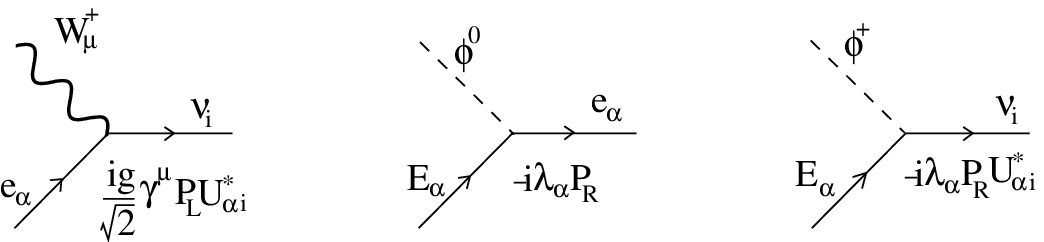}}
\caption{(Some of) the standard model Feynman rules. Here, $\lambda$ are charged lepton Yukawa couplings, $\lambda_{\alpha}=m_{\alpha}/v$, $U_{\alpha i}$ are the elements of the leptonic mixing matrix, and $P_{L,R}$ are, respectively, left and right-handed chirality projection operators. 
\label{fig:feynSM}}
\scalebox{1.1}{\includegraphics{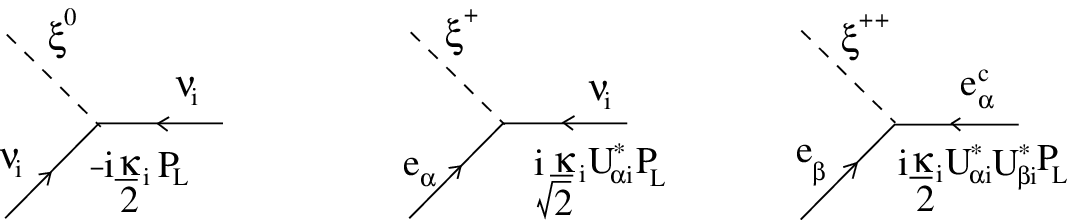}}
\caption{(Some of) the triplet Higgs boson Feynman rules. $U_{\alpha i}$ are the elements of the leptonic mixing matrix, and $P_{L,R}$ are, respectively, left and right-handed chirality projection operators.
\label{fig:feynLxiL}}
\scalebox{1.1}{\includegraphics{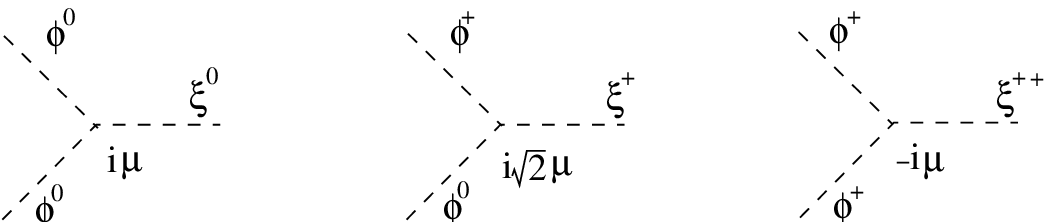}}
\caption{The triplet Higgs boson explicitly $L_\#$ violating Feynman rules. 
\label{fig:feynVmu}}
\end{center}
\end{figure}

Expanding the $SU(2)_L$ structure of Eq.~(\ref{eq:Vmu})
\beq
V_\mu = -\mu \left( {\xi^0}^* \phi^0 \phi^0 + \sqrt{2} \xi^- \phi^0 \phi^+ - \xi^{--}\phi^+ \phi^+ \right) + {\rm h.c.} \ ,
\label{eq:Vmuexp} 
\eeq
which describes the couplings between the Higgs doublet and triplet (including Goldstone bosons).
The Feynman rules for this model are depicted in Figs.~\ref{fig:feynSM},~\ref{fig:feynLxiL}~and~\ref{fig:feynVmu}.

\subsection{Brief Phenomenology of the Higgs Triplet Model}

In this subsection, we briefly discuss constraints and signatures of the triplet Higgs model in the case $M\lesssim 1$~TeV,  concentrating on qualitatively estimating the current experimental constraints on the dimensionless couplings $\kappa$.
We note that, given that constraints from precision electroweak observables are proportional to 
$u$ (of order the neutrino masses), these are, therefore not significant~\cite{Gunion:1989ci}, and will not be explored.

\subsubsection{Charged Lepton Flavor Violation: $\mu\rightarrow e \gamma$, $\mu\rightarrow e$ Conversion in Nuclei, $\mu\rightarrow e e e$}
Here we estimate the size of $\mu \rightarrow e$ flavor changing neutral current processes. These depend, in
general, on details of the neutrino mass spectrum and mixing angles, and   
a numerical study of these processes taking the best fit values to the solar and atmospheric 
oscillation data has been carried out in Ref.~\cite{Kakizaki:2003jk}.\footnote{In that study, 
$\mu = 25~{\rm eV}$ and $M = 200~{\rm GeV}$ were used.}

\begin{figure}
\begin{center}
\scalebox{1.1}{\includegraphics{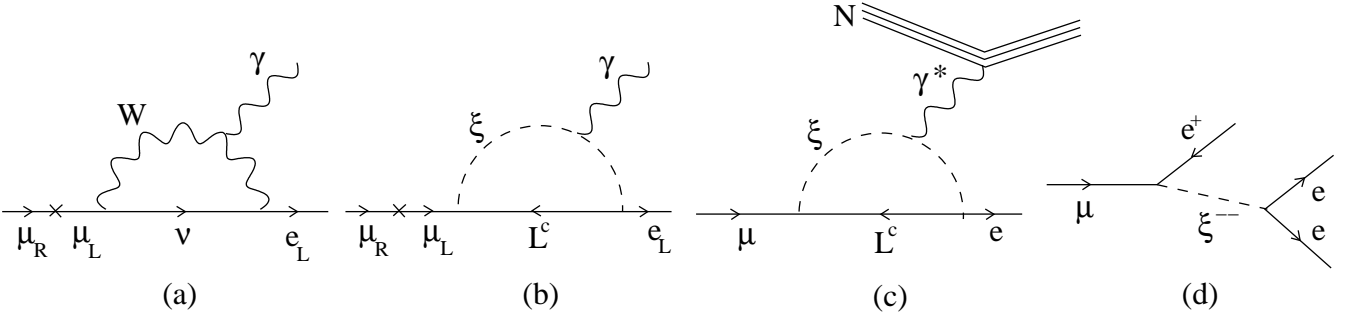}}
\caption{One-loop diagrams that mediate $\mu\rightarrow e\gamma$ in the SM and in the triplet Higgs boson model ((a) and (b), respectively), one-loop diagram that mediates $\mu\to e$ conversion in nuclei in the triplet Higgs boson model (c) and tree level doubly charged Higss exchange contribution to $\mu\to eee$. See text for details. 
\label{fig:mutoe}}
\end{center}
\end{figure}
The experimental limit on the branching ratio~(BR) is  
${\rm BR}(\mu\rightarrow e \gamma) < 1.2\times 10^{-11}$ at 90\,\% C.L.~\cite{Brooks:1999pu}.
The one-loop SM contribution shown in Fig.~\ref{fig:mutoe}(a) is tiny, due to GIM suppression \cite{meg_sm}. On the other hand, the triplet Higgs one-loop contribution depicted in Fig.~\ref{fig:mutoe}(b) 
is no longer GIM suppressed, and is given by
\beq
{\rm BR}^{(\xi)}(\mu\rightarrow e \gamma) \approx \frac{e^2}{(16\pi^2)^2} \frac{\left| (U\kappa^2 U^\dagger)_{e\mu} \right|^2}{g^4} \left(\frac{m_W}{M}\right)^4 \ .
\eeq
The BR depends on $(U\kappa^2 U^\dagger)_{e\mu} = (\bar\kappa^\dagger \bar\kappa)_{e\mu}$, and, 
therefore, on the details of the neutrino mass spectrum and mixing angles.
For $M\sim 1~{\rm TeV}$, we find from the experimental limit that 
$\left|(\bar\kappa^\dagger \bar\kappa)_{e\mu}\right| \lesssim 0.1$ has to be satisfied.

The experimental upper bound on the  $\mu\rightarrow e$ conversion rate in titanium normalized to the muon capture rate 
is $4.3\times 10^{-12}$~\cite{Dohmen:1993mp}. 
One triplet Higgs boson contribution is depicted in Fig.~\ref{fig:mutoe}(c). 
$\mu\rightarrow e$ conversion has the helicity flip contribution similar to 
$\mu\rightarrow e \gamma$, with the photon connected to the nucleus, 
which brings in a suppression of $\alpha$.  
However, since the photon can be off-shell, there are additional non-helicity-suppressed 
contributions~\cite{Marciano:1977cj} that overcome the $\alpha$ suppression, and the 
$\mu\rightarrow e$ conversion rate becomes comparable to that of
$\mu\rightarrow e \gamma$~\cite{Kakizaki:2003jk}.

The experimental limit on ${\rm BR}(\mu\rightarrow e e e)$ is ${\rm BR}(\mu\rightarrow e e e)<10^{-12}$~\cite{Bellgardt:1987du}.
The one-loop SM contribution is again tiny due to GIM suppression \cite{meg_sm}.
The doubly-charged triplet Higgs boson mediates this decay at tree level, as depicted in 
Fig.~\ref{fig:mutoe}(d), and we estimate that
\beq
{\rm BR}^{(\xi)}(\mu\rightarrow e e e) \sim \frac{\left|(U^*\kappa U^\dagger)_{e\mu} (U^*\kappa U^\dagger)_{ee}^* \right|^2}{g^4} \left(\frac{m_W^2}{M^2}\right)^2 \ .
\eeq
The BR depends on $(U^*\kappa U^\dagger)_{ee,e\mu}= \bar\kappa_{ee,e\mu}$, and, from the 
experimental limit, for $M\sim 1~{\rm TeV}$, we obtain the bound, 
$\left|\bar\kappa_{e\mu} \bar\kappa_{ee}^* \right| \lesssim 10^{-4}$. 
 
\subsubsection{Muonium-Antimuonium Oscillations}
\begin{figure}
\begin{center}
\scalebox{1.1}{\includegraphics{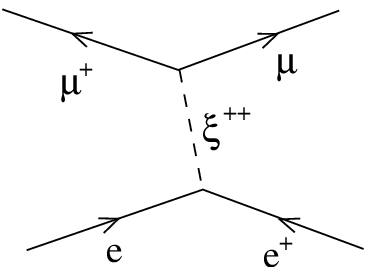}}
\hspace{2cm}
\scalebox{1.1}{\includegraphics{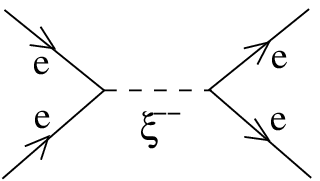}}
\caption{Tree level doubly charged Higss exchange contribution to muonium--antimuonium oscillations (left) and M\o ller scattering (right). See text for details.  
\label{fig:muamuosc}}
\end{center}
\end{figure}
The current experimental limit on muonium-antimuonium oscillation (see for example 
Ref.~\cite{Clark:2003tv}), given in terms of the dimension-six effective
interaction coefficient $G_C$, is \cite{Willmann:1998gd}
\beq
R_g \equiv \frac{G_C}{G_F} < 0.003 \ .
\eeq
The doubly-charged triplet Higgs boson mediates this oscillation at tree-level,
as depicted in Fig.~\ref{fig:muamuosc}(left), and we estimate
\beq
G_C \sim \frac{(U\kappa U^T)_{\mu\mu} (U^*\kappa U^\dagger)_{ee}}{M^2} \ .
\eeq
Since $U\kappa U^T = \bar\kappa^\dagger$, from the experimental limit we derive the bound,
$|\bar\kappa_{\mu\mu}^* \bar\kappa_{ee}|/M^2 \lesssim 10^{-8}~{\rm GeV}^{-2}$, and for 
$M\sim 1~{\rm TeV}$ this implies $|\bar\kappa_{\mu\mu}^* \bar\kappa_{ee}| \lesssim 10^{-2}$. 

\subsubsection{M\o ller Scattering}

The current experimental measurement of the parity-violating asymmetry in 
polarized M\o ller scattering, $e^- e^- \rightarrow e^- e^-$ (see for example 
Refs.~\cite{Puhala:1981qe,Czarnecki:2000ic}), agrees with SM predictions.
This implies a bound on the scale $\Lambda$ of any new dimension-six four-electron contact 
interaction, which is (for an operator coefficient of $2\pi$), 
$\Lambda \gtrsim 7.2~{\rm TeV}$~\cite{Anthony:2003ub}. 
The doubly-charged triplet Higgs contributes to M\o ller
scattering at tree level, as we show in Fig.~\ref{fig:muamuosc}(right),
and the above limit on $\Lambda$ translates into the bound,
\beq
\frac{|(U\kappa U^T)_{ee}|^2}{M^2} \lesssim \frac{2\pi}{(7.2~{\rm TeV})^2} \ .
\eeq
From this, using $U\kappa U^T = \bar\kappa^\dagger$, we derive the bound 
$|\bar\kappa_{ee}|^2/M^2 \lesssim 1/(2.8~{\rm TeV})^2$, and for $M\sim 1~{\rm TeV}$, we get
$|\bar\kappa_{ee}|^2 \lesssim 0.1$. 

\subsubsection{Neutrinoless Double-Beta Decay ($0\nu\beta\beta$)}
\begin{figure}
\begin{center}
\scalebox{1.1}{\includegraphics{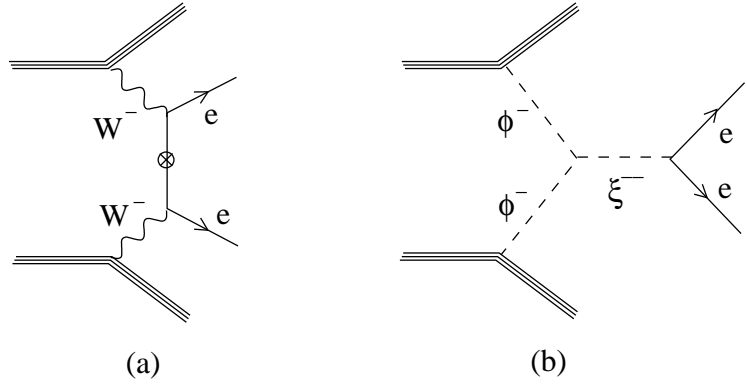}}
\caption{Tree level doubly charged Higss exchange contribution to neutrinoless double-beta decay. See text for details. 
\label{fig:zeronubb}}
\end{center}
\end{figure}
Observing a nonzero rate for $0\nu\beta\beta$ unambiguously implies that $L_\#$ is broken
and that neutrinos are Majorana particles. The effective neutrino Majorana mass matrix in
Eq.~(\ref{eq:Lnumeff}) leads to $0\nu\beta\beta$ as shown in Fig.~\ref{fig:zeronubb}(a), 
and has been analyzed exhaustively in the literature (see for example
Ref.~\cite{Elliott:2004hr}).  
The $0\nu\beta\beta$ decay amplitude arising from such an effective neutrino mass is 
\beq
{\cal A}^{(m_\nu)} \propto g^4 \frac{(U m U^T)_{ee}}{m_W^4 q^2} \ ,
\label{sm_0nubb}
\eeq
with $q^2 \approx (50~{\rm MeV})^2$, and we have $(U m U^T)_{ee} = \bar{m}^*_{ee}$. 
The current experimental limit implies the bound 
$|\bar{m}_{ee}| \lesssim 0.38~{\rm eV}$ (see, for example, \cite{global_anal}).  
In addition to this, the doubly-charged Higgs triplet contributes as shown in
Fig.~\ref{fig:zeronubb}(b), and is given by
\beq
{\cal A}^{(\xi)} \propto \frac{\lambda_u \lambda_d (U\kappa U^T)_{ee} \mu^*}{m_W^4 M^2} \ ,
\eeq
where $\lambda_u,\lambda_d\sim 10^{-5}$ are the $u,d$ quark Yukawa couplings, and
we note that $(U\kappa U^T)_{ee} = \bar\kappa_{ee}^*$. $\bar{\kappa}_{ee}^* \mu^*/M^2=\bar{m}_{ee}^*/v^2$, so that this contribution is severely suppressed with respect to the massive Majorana neutrino one, Eq.~(\ref{sm_0nubb}).

 \end{document}